# High efficiency deep-UV emission at 219 nm from ultrathin MBE GaN/AlN quantum heterostructures

SM Islam[1], Vladimir Protasenko[1], Kevin Lee[1], Sergei Rouvimov[2], Jai Verma[3*], Huili (Grace) Xing[1,4], and Debdeep Jena[1,4]

[1]*Department of Electrical & Computer Engineering, Cornell University, NY 14853, USA*
[2]*Notre Dame Integrated Imaging Facility, University of Notre Dame, IN 46556, USA*
[3]*Department of Electrical Engineering, University of Notre Dame, IN 46556, USA*
[4]*Department of Materials Science and Engineering, Cornell University, NY 14853, USA*
**Presently at Intel Corporation*

Deep ultraviolet (UV) optical emission below 250 nm (~5 eV) in semiconductors is traditionally obtained from high aluminum containing AlGaN alloy quantum wells. It is shown here that high-quality epitaxial ultrathin binary GaN quantum disks embedded in an AlN matrix can produce efficient optical emission in the 219-235 nm (~5.7 to 5.3 eV) spectral range, far above the bulk bandgap (3.4 eV) of GaN. The quantum confinement energy in these heterostructures is larger than the bandgaps of traditional semiconductors, made possible by the large band offsets. These MBE-grown extreme quantum-confinement GaN/AlN heterostructures exhibit internal quantum efficiency as high as 40% at wavelengths as short as 219 nm. These observations, together with the ability to engineer the interband optical matrix elements to control the direction of photon emission in such new binary quantum disk active regions offers unique advantages over alloy AlGaN quantum well counterparts for the realization of deep-UV light-emitting diodes and lasers.

The ruggedness, portability, high-efficiency, and microfabrication benefits of solid-state semiconductor light sources over conventional lamps became clear in the last decade for visible wavelengths in the solid-state lighting revolution, and gave birth to several new applications. A similar revolution is expected in the deep-UV spectrum. Semiconductor light sources such as Light-Emitting Diodes (LEDs) and Lasers in the deep ultraviolet (UV) spectrum have versatile applications in water and air purification, in healthcare applications of bio-photonic diagnostics and sterilization, in food preservation, in security and environmental monitoring and in industrial curing. The semiconductor material substrate of choice for deep-UV photonic devices is direct-bandgap AlN with an energy bandgap of ~6.2 eV (200 nm), and the active regions where photons are produced are various ternary compositional alloys of AlN with GaN of bandgap ~3.4 eV (365 nm).

For deep-UV LEDs, quantum well active regions composed of AlGaN have been used to push the interband optical transition to high energies [1-5]. The internal quantum efficiency in high Al containing AlGaN Quantum Wells (QWs)/barrier structures is limited by the quantum confined stark effect (QCSE) [6-8], edge emission due to valence band structure re-ordering [9-11], combined with material defect (e.g. dislocation) induced non-radiative recombination. Compositional fluctuations of Al and Ga concentrations in ternary AlGaN alloy layers degrade efficient optical emission in the deep-UV range [12], and together with the other effects degrade the LED efficiency.

Distinct from the alloy AlGaN layers, deep-UV emission down to 224 nm has been achieved in binary GaN/AlN heterostructures [13-17]. As a significant advantage, the polarization of the emitted photons in ultrathin GaN QWs and quantum dots/disks (QDs) is perpendicular to the c axis, making them propagate parallel to the c-axis [9,11]; this surface emission property is highly favorable for light extraction.

We recently demonstrated deep UV LEDs [18-20] emitting as short as 232 nm by incorporating 2 monolayer (ML) thick GaN QDs in AlN barriers. As the height of the QD reduces and the oscillator strength increases [21], the radiative lifetime decreases significantly, increasing the internal quantum efficiency. Shortening the emission wavelength even deeper below 230 nm by utilizing GaN QDs embedded in AlN barriers will further enable applications in sensing and toxic gas detection applications. Tunable sub-230 nm deep-UV emission was demonstrated by Molecular Beam Epitaxy (MBE) growth of 2 ML GaN QDs using a modified Stranski-Krastanov (m-SK) growth method [22]. The m-SK technique uses thermal annealing of the 2 ML GaN quantum well structure sandwiched between AlN barriers. In this letter, we present an *alternative* approach to realize tunable sub-230 nm emission with higher internal quantum efficiency using SK growth of 2 ML GaN QD structures by MBE. Unlike the earlier work based on m-SK method, control of the emission wavelength is achieved by changing the Ga/N ratio in a Nitrogen-rich growth regime that ensures dot/disk formation. This SK growth method for GaN QDs enables us to achieve shorter emission wavelength (219 nm with SK compared to 222 nm with m-SK) with higher internal quantum efficiency (40% with SK compared to 36% with m-SK).

The samples studied were grown using a Veeco Gen-930 plasma assisted MBE system on 1 um thick AlN templates on Sapphire of threading dislocation density ~$10^{10}$ cm$^{-2}$. After standard solvent cleaning the samples were loaded in the MBE chamber and outgassed for 7 hrs at 200°C followed by ~2 hrs at 450°C. Prior to the epitaxy of heterostructures discussed in this work, calibration growths were performed to identify the most suitable AlN barrier thickness, GaN annealing time and the growth rate for the most intense deep-UV emission. Based on these studies, three samples containing [2 ML GaN/4 nm AlN barrier] heterostructures shown in Fig. 1(a) were grown under conditions indicated in Fig 1(b). The substrate thermocouple temperature was 730ºC throughout the growths, with an Al flux of $1x10^{-7}$ Torr beam equivalent pressure. The RF plasma power of 275 W with 1.2 sccm flow led to an effective nitrogen BEP of $1x10^{-7}$ Torr (at a chamber pressure of ~$2x10^{-5}$ Torr). The Ga flux was varied for the three samples to change the size of the QDs to tune the UV emission wavelength.

The active region contained 10 periods of the GaN/AlN heterostructure grown on a 30 nm AlN buffer layer as shown in Fig 1(a). The 4 nm AlN barriers were grown using migration enhanced epitaxy (MEE) [23] to ensure a smooth heterointerface between AlN and GaN by avoiding excess Al. Because of the strong preference of Al incorporation over Ga, an Al-free surface is necessary to incorporate ultra-thin GaN QDs in the AlN matrix. RHEED oscillations [24] were used to precisely determine and control the growth rate at 0.29 ML/s and to maintain desired stoichiometry for AlN and GaN. The 2 ML GaN QDs were grown in N-rich conditions by opening the Ga and N shutters simultaneously for 7s ensuring SK growth mode. A growth interrupt (anneal) of 18s was introduced after each GaN layer deposition to assist island formation by layer decomposition. The Ga/N ratio was varied from 0.88 (A)→0.75 (B)→0.6 (C) for the three samples. Fig. 1b shows the flux-time diagram for 1 complete cycle of the quantum dot superlattice active region.

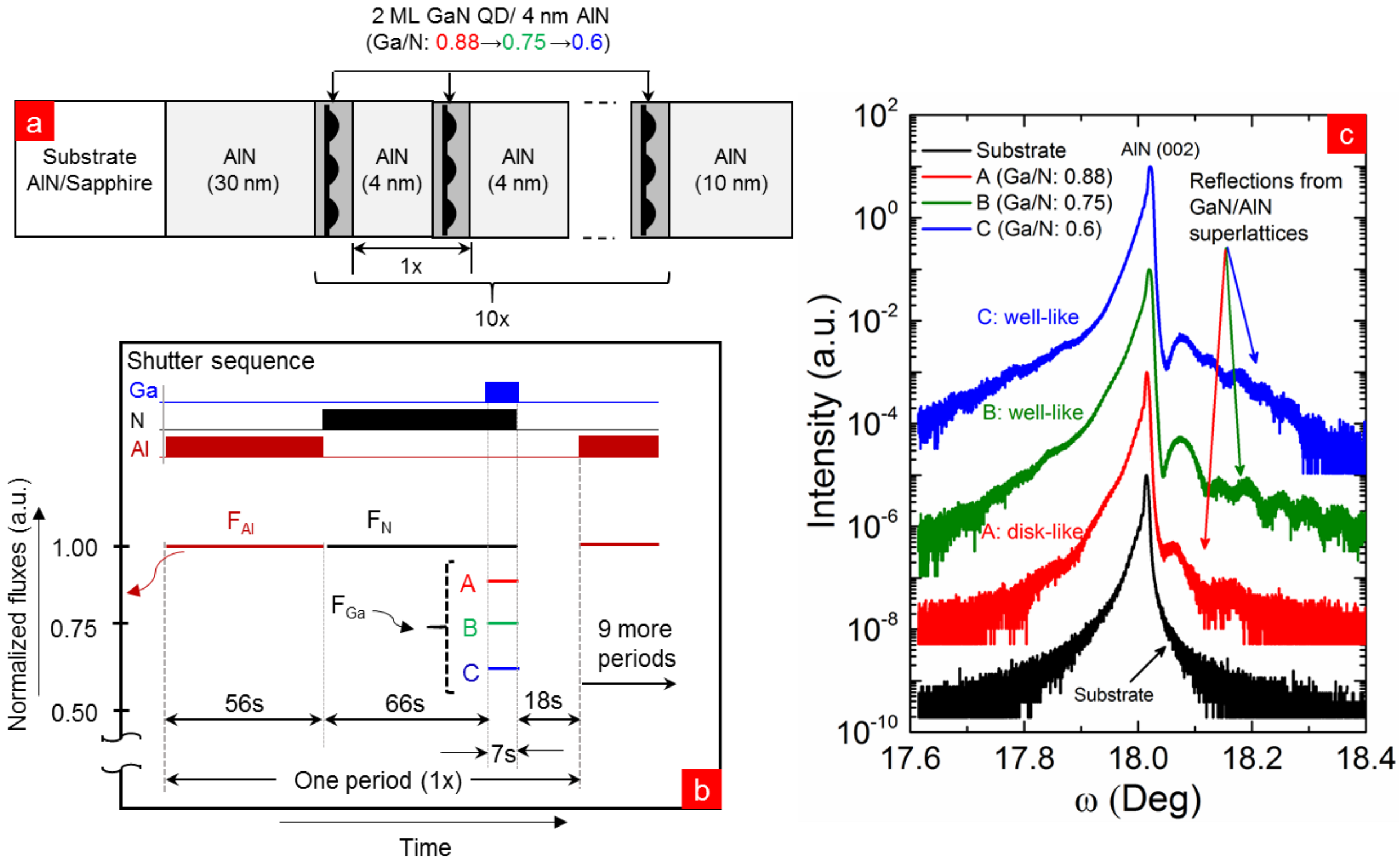

Fig. 1: a) Schematic of the structure with 10 periods of ultra-thin GaN QDs in AlN barriers, b) MBE growth diagram showing shutter sequence and relative III-V fluxes for 1 period of the growth, c) HR-XRD ω-2θ scans showing signature of QD formation for sample A.

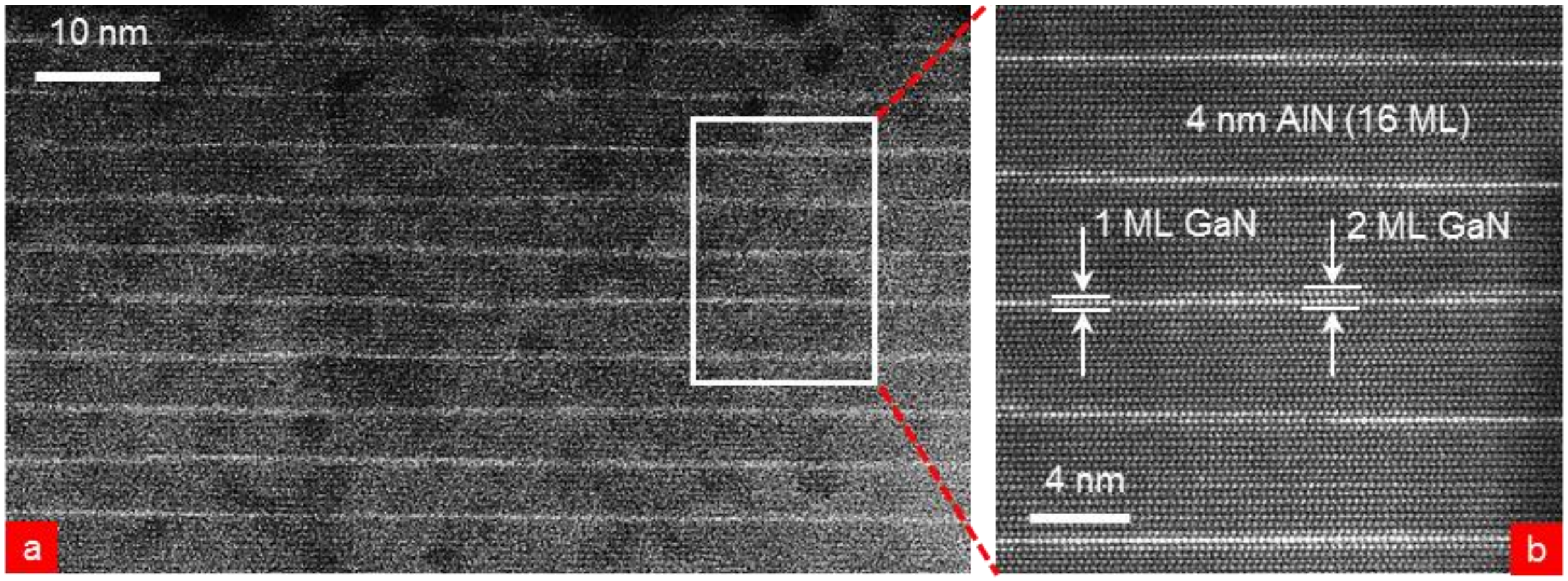

Fig. 2: Z-contrast STEM images for sample A, (a) large area scan showing uniform distribution of the 10 periods of GaN/AlN heterostructures, (b) zoomed-in image showing presence of 1 ML and 2 ML GaN QDs separated by 4nm AlN.

Fig. 1(c) shows the measured triple-axis X-Ray Diffraction (002) ω-2θ spectrum. The AlN peak in all samples is similar to the control substrate which was 1 μm AlN on 430 μm sapphire. All three GaN/AlN UV emitter samples show satellite fringes due to reflections from the heterointerfaces confirming periodic GaN/AlN heterostructures. The fringes for samples B and C were more prominent than sample A, suggesting sample A to be more disk-like and B/C to be more well-like quantum structures [25-26]. QD-like features cause diffused scattering of the XRD beam preventing well defined satellite peaks [25]. As the Ga/N ratio is reduced from 0.88 to 0.6, reflection from the well-like GaN planes produce constructive interference [26] and strong XRD fringes.

Fig. 2 (a & b) show Z-contrast Scanning Transmission-Electron Microscope (STEM) images of Sample A which was grown with a 0.88 Ga/N ratio. Sharp heterointerfaces were observed between GaN and AlN suggesting no intermixing to form AlGaN alloy. The existence of 1-2 ML GaN QDs is clearly indicated in Fig. 2b. Also in some places the GaN layer was completely absent. The image indicates that by growing GaN under N-rich conditions, the effective QD thickness can be varied between 0-2 monolayers.

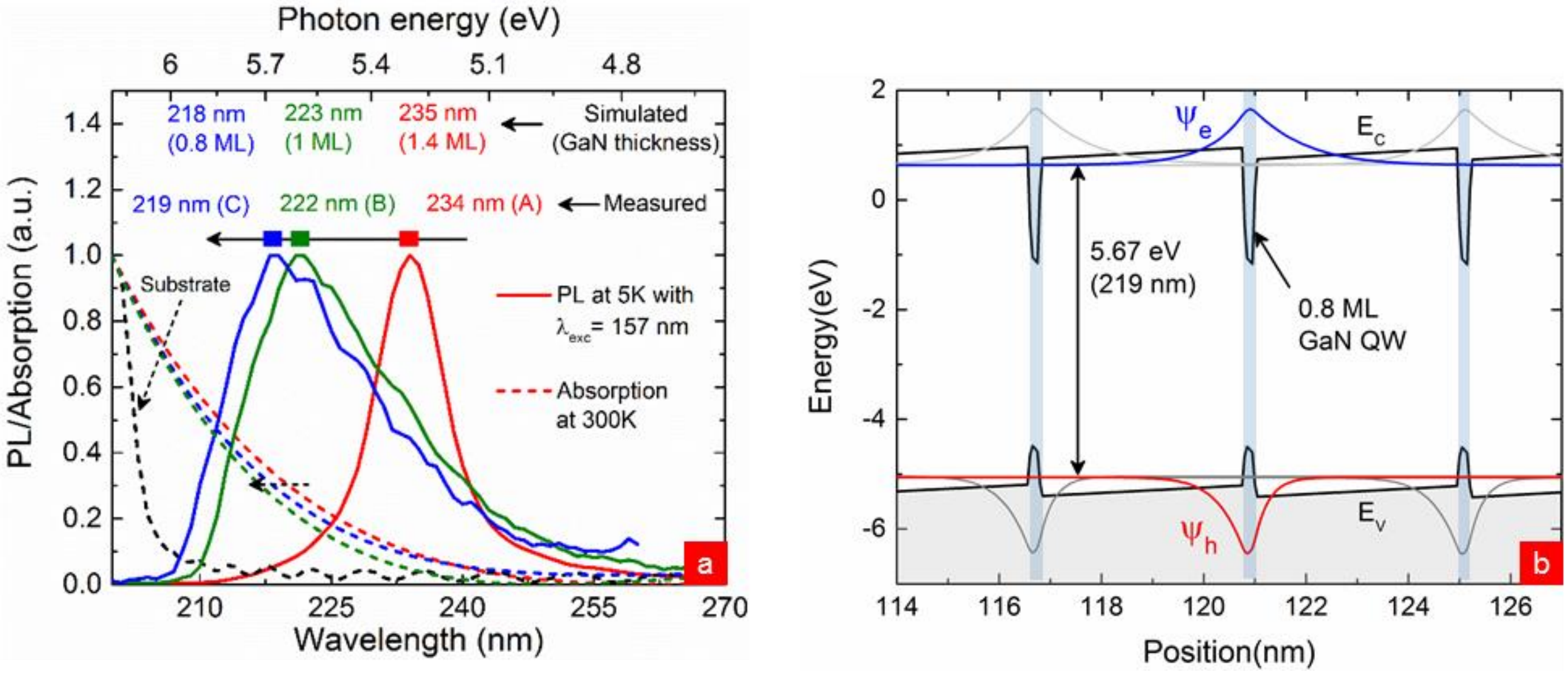

Fig. 3: a) Normalized measured photoluminescence (5 K) and absorption (300 K) spectra showing tunable deep-UV emission down to 219 nm. The average thickness of GaN layers were extracted from simulation, b) simulated energy-band diagram for 219 nm emission.

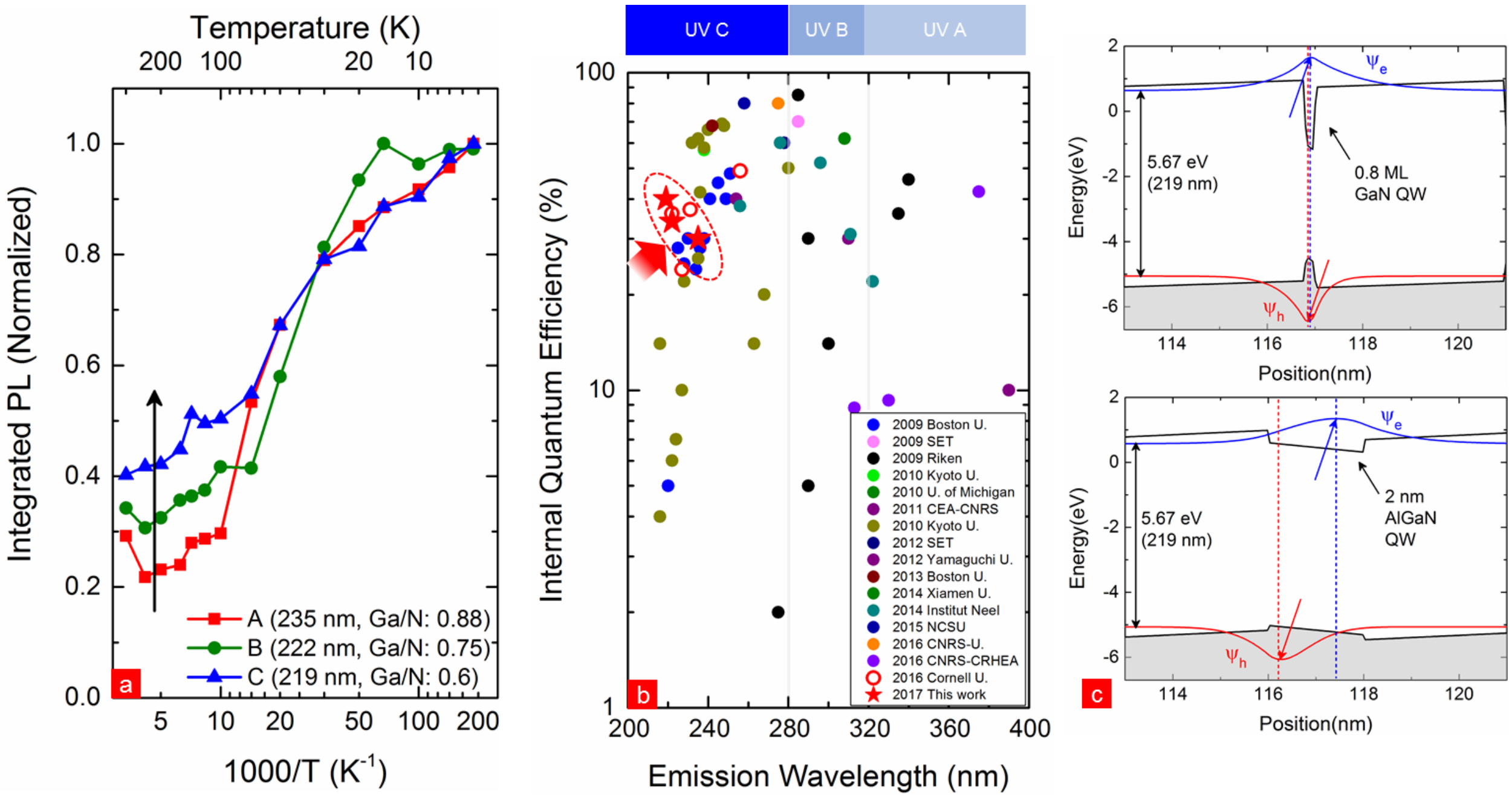

Fig. 4: a) Estimation of IQE from temperature dependent integrated PL spectra, b) comparison of measured IQE with AlGaN based QWs, c) enhancement of IQE using GaN QWs due to reduced quantum-confined-stark-effect.

Fig 3(a) shows the measured 5K photoluminescence (PL) and 300K absorption (Abs) spectra. A pulsed Excimer laser (λ= 157 nm, 10 ns pulse duration, 600 μJ/pulse energy, 157 Hz repetition rate) was used for PL excitation. Sample A with the highest Ga/N ratio of 0.88 shows a peak PL emission at 234 nm, sample B (Ga/N: 0.75) at 222 nm, and sample C with the lowest

Ga/N ratio of 0.6 exhibits a 219 nm PL peak at 5 K. This is the *highest* reported emission energy to-date using GaN as the light emitting material. Temperature-dependent PL measurements showed that at 300 K the PL peak wavelength red-shifted by ~4 nm consistent with a Varshni trend. The measured PL peak positions were compared with a Schrodinger-Poisson simulation using SiLENSe software; an example energy-band diagram is shown in Fig 3(b). The effective GaN thicknesses for the three samples to match the peak PL wavelength with experiment was ~1.4 ML (A), 1 ML (B) and 0.8 ML (C). Thus the emission wavelength can be reproducibly controlled by choosing the appropriate Ga/N ratio. The absolute integrated PL intensity was found to decrease at the shorter wavelengths with lowering the Ga/N ratio. This is expected because of the reduction of the total GaN volume at higher N-rich growth conditions. The high electron-hole overlap seen in Fig. 3b calculated for the 0.8 ML effective GaN thickness indicates robustness to quantum-confined Stark effect, and leads to an interband transition wavelength of 219 nm (5.67 eV) for deep-UV emission, consistent with the experimental observation.

The PL linewidth is a measure of the thickness fluctuations of the GaN layers. For example, the average thickness of the GaN layers for each of the samples in Fig. 3a are 1.4 ML (A), 1 ML (B) and 0.8 ML (C) as mentioned before. Based on the STEM image in Fig. 2(b), the thickest GaN layer region in the samples is 2 ML. Therefore the thickness fluctuation (Δz) for the three samples are ~0.6 ML (A), ~1 ML (B) and ~1.2 ML (C) respectively. The broadening is estimated using the formula $(dE_0/dz)\cdot\Delta z$ where $dE_0/dz$ is the differential change of the Eigenvalue energies calculated at the effective GaN thicknesses for the three samples. The simulation tool SiLENSe is used to calculate $dE_0/dz$ for each of the samples. Based on this analysis, the calculated broadening due to the thickness fluctuation are 16 nm (A), 30 nm (B) and 36 nm (C) which are higher than the measured values of 9 nm (A), 20 nm (B) and 19 nm (C). This qualitative agreement shows the correct trend and order of magnitude, but to obtain quantitative agreement it is necessary to incorporate the size variations with full-bandstructure models [17] that is not attempted here. The measured 300 K absorption spectra shown in Fig. 3a. Because the three samples contain relatively small absorbing volumes estimated as 10 periods of 2 ML GaN, resulting in <5 nm of total effective thickness, the absorption edge is not sharp enough for a quantitative absorption edge. Nevertheless, the absorption edges are consistent with the measured PL emission spectra.

Fig. 4a shows the temperature-dependent integrated PL spectra which is indicative of the internal quantum efficiency (IQE). The IQE estimated from the 300K/5K ratio was 29.2% for sample A grown at 0.88 Ga/N ratio. The IQE went up to 40.2% for sample C with 0.6 Ga/N ratio. The increase in IQE is explained by the reduction of the QCSE. Fig. 4c (top & bottom) compares the energy band diagrams and electronic states of an ultrathin GaN QW/QD with a corresponding AlGaN QW with the *same* photon emission energy. The significantly higher overlap of the electron/hole wavefunctions indicate a higher oscillator strength, which should translate to a higher IQE.

Fig. 4b shows the measured IQE for the heterostructures reported in this work in comparison to those reported in the literature over the range of 200-400 nm [27-43]. To increase the photon energy in traditional AlGaN based thick (>2 nm) QW active regions, the Al-content in the AlGaN is typically increased. The high QCSE and lower band offset induced carrier leakage into the barriers reduces the IQE of AlGaN active regions. Beyond an Al content of 40% [9], the optical matrix element leads to edge-emission from AlGaN QWs. On the contrary, for the GaN based structures shown in this work, the IQE is found to *increase* at shorter wavelength emission. This is achieved because the reduction of the GaN QW/QD thickness enhances the overlap integral, reversing the trend of decreasing IQE at shorter wavelengths. Furthermore, the emission matrix element is converted to surface-emission [13], which is highly desirable for light extraction.

In conclusion, ultrathin (1-2ML) GaN dots/disks in AlN matrix were demonstrated to be capable of controllable emission in the 219-235 nm range by engineering the quantum confinement during the MBE growth, showing the shortest wavelength emission till date at 219 nm (~5.7 eV) from binary GaN active regions. An IQE as high as 40% at an emission wavelength of 219 nm was measured, which is more than 2X higher than the highest prior reported AlGaN QW based heterostructures at comparable short deep-UV wavelengths. Together with the reduced QCSE and the surface emission, the use of the extreme quantum confinement binary GaN QW/QD active regions offer a compelling new approach for efficient deep UV light emitters.

This work was supported in part by a NSF DMREF grant #1534303, and an AFOSR grant monitored by Dr. Ken Goretta.